\documentstyle[11pt]{article}
\begin{document}
\begin{center}
{\large Proceedings of the European Workshop on Heavy
Element Nucleosynthesis, Budapest, March 9-11, 1994,
in press}
\end{center}
\begin{center}
{\Large \bf DIRECT CAPTURE AT LOW ENERGIES}
\end{center}
\begin{center}
W.~BALOGH, R.~BIEBER, H.~OBERHUMMER
(Institut f\"ur Kernphysik, TU Wien, Wiedner Hauptstr.~8--10,
 A--1040 Vienna, Austria)
\end{center}
\begin{center}
T.~RAUSCHER, K.-L. KRATZ
(Institut f\"ur Kernchemie, Univ.~Mainz, Fritz--Strassmann Weg 2,
 D--55099 Mainz, Germany)
\end{center}
\begin{center}
P.~MOHR, G. STAUDT
(Physikalisches Institut, Univ.~T\"ubingen, Auf der Morgenstelle 14,
D--72076 T\"ubingen, Germany)
\end{center}
\begin{center}
M.M.~SHARMA
(Max--Planck--Institut fŸr Astrophysik, Postfach 1523,
Karl--Schwarzschild--Str.~1
D--85740 Garching, Germany)
\end{center}
{\bf Abstract:} The importance of direct capture for
(n,$\gamma$)--reactions on intermediate-- and heavy--mass target nuclei
occuring in the s-- and r--process is investigated.
It is shown that the direct mechanism is non--negligible for magic
and neutron rich
target nuclei. For some double magic and neutron rich
nuclei in the r--process direct capture is even the dominant reaction
mechanism.
\newpage
\section{Introduction}

The reaction mechanism of radiative capture can be
classified by two extremes: the reaction proceeds
as a multi--step process (compound--nucleus reaction: CN)
or in a single step (Direct Capture: DC).
In the last years it was realized that for light nuclei
DC is often the dominating reaction mechanism in
astrophysically relevant nuclear processes.
The DC--method together with the folding procedure has been
applied sucessfully by our group to radiative capture in primordial
nucleosynthesis ($^{2}$H($\alpha$,$\gamma$)$^{6}$Li \cite{moh1},
$^{3}$H($\alpha$,$\gamma$)$^{7}$Li \cite{moh2},
$^{7}$Li(n,$\gamma$)$^{8}$Li \cite{kra1}),
the pp--chain ($^{3}$He($\alpha$,$\gamma$)$^{7}$Be \cite{moh2},
$^{7}$Be(p,$\gamma$)$^{8}$B \cite{kra1}),
CNO--cycle ($^{13}$N(p,$\gamma$)$^{14}$O \cite{dec})
and helium burning
($^{8}$Be($\alpha$,$\gamma$)$^{12}$C \cite{obe1}).

In this work we want
to investigate the importance of the DC mechanism for
(n,$\gamma$)--reactions on intermediate-- and heavy--mass target nuclei
occuring in the s-- and r--process.
In Section 2 we introduce the DC formalism
and the folding procedure. In Section 3 the relativistic
mean field theory (RMFT) and its application
to neutron rich nuclei will be discussed.
Finally in Section 4 we compare the influence of DC to
radiative capture cross sections in the s-- and r--process.
\section{Direct capture and folding procedure}
The theoretical cross section $\sigma^{\rm th}$ is obtained from the
DC cross section $\sigma^{\rm DC}$ given by \cite{kim},
\cite{obe2}, \cite{moh2}
\begin{equation} \label{1}
\sigma^{\rm th} = \sum_{i} \: C_{i}^{2} S_{i}\sigma^{\rm DC}_{i} \quad .
\end{equation}
The sum extends over all possible final states (ground state and excited
states)
in the final nuclei. The isospin Clebsch--Gordan coefficients
and spectroscopic factors are $C_{i}$ and $S_{i}$,
repectively. The DC cross sections $\sigma^{\rm DC}_{i}$ are essentially
determined by the overlap of the scattering wave function
in the entrance channel, the bound--state wave function
in the exit channel and the multipole transition--operator.
For the
computation of the DC cross section we used the direct--capture code TEDCA
\cite{kra2}.

The folding procedure is used for calculating the
nucleon--nucleus potentials in order to describe the
elastic scattering data and the bound states. This method was already applied
successfully in describing many nucleon--nucleus systems.
In the folding approach the nuclear density
$\rho_{A}$
is derived from experimental charge distributions \cite{deV} and folded
with an energy and density dependent NN interaction
$v_{\rm eff}$ \cite{kob}:
\begin{equation} \label{2}
V(R) = \lambda V_{\rm F}(R) = \lambda \int
\rho_{A} (\vec{r}) v_{\rm eff}(E,\rho_A,|\vec{R}  - \vec{r}|)
d\vec{r} \quad .
\end{equation}
with $\vec{R}$ being the separation of the centers of mass of the two colliding
nuclei. The normalization factor $\lambda$ is adjusted to elastic scattering
data
and to bound-- and
resonant--state energies. The potential obtained in
this way ensures the correct behavior of the wave functions in the nuclear
exterior.
At the low energies considered in nucleosynthesis the
imaginary parts of the optical potentials are small.
The folding potentials of
Eq.~\ref{2} were determined with the help of the
computer code DFOLD \cite{abe}.
\section{Application of RMFT to neutron rich nuclei}
The RMFT describes the nucleus as a system of Dirac nucleons interacting via
various meson fields. In the last few years this theory has turned out to be a
very successful tool for the description of many nuclear properties (for
example binding energies and charge radii for stable isotopes) \cite{gam}.
The RMFT is built upon two main approximations: the
mean--field approximation and the no--sea approximation. The mean--field
approximation removes all quantum fluctuations of the meson fields
and uses their expectation values. This approach cuts down all many body
effects because
the nucleons move as independent particles in the meson fields. The no--sea
approach neglects all contributions from antiparticles. Only negative
energy states are taken into account. One great advantage of the RMFT is that
the spin--orbit interaction is described in a
proper way without any additional parameters.

The theory starts with an effective interaction of Dirac nucleons with
mesons and the electromagnetic field. We work with
the ($\sigma\omega\rho$)--model \cite{ser}.
The ${\sigma}$--meson mediates the medium range attraction between the
nucleons.  The isoscalar vector mesons ${\omega}$ cause a short--range
repulsion.
The contribution of the ${\rho}$--mesons is important for neutron--
and proton--rich nuclei. There are six parameters which are usually
obtained by fits to finite nuclear properties:
each coupling constant of the meson fields with the nucleons $g_{\sigma}$,
$g_{\omega}$ and $g_{\rho}$, the constants of the nonlinear
${\sigma}$--potential $g_2$,
$g_3$, and the mass of the ${\sigma}$--meson. The other meson masses are
given empirically.

In our calculations we have used the parameter sets NLSH \cite{sha1},
\cite{sha2}
and NL1 \cite{rei}. The parameter set NL1 is fitted to the ground
state binding energies and charge radii of spherical nuclei. This set
overpredicts the neutron radii of neutron rich nuclei. The set NLSH has been
fitted
not only to binding energies and charge radii, but also to the neutron
radii of several spherical nuclei.
\begin{figure}
\vspace{18cm}
\caption{Upper part: Binding energies of Sn--isotopes calculated
with RMFT using the parameter set NLSH.
Lower part: Neutron--skin thickness of Sn--isotopes calculated
with RMFT using the parameter sets NL1 (broken curve) and NLSH
(solid curve) compared with experimental data [17].}
\end{figure}
The binding energies of Sn--isotopes as a function of
mass number show a kink at $A=132$ (upper part of Fig.~1)
signifying the shell effect at the magic neutron number $N=82$.
The neutron--skin thickness of the Sn--isotopes also show
this kink by about $0.25 (N-Z)/A$ (lower part of Fig.~2).
The values for the neutron--skin thickness calculated with the parameter set
NLSH
are about 20\,$\%$ less
than the values determined with NL1. The results of NLSH for
the binding energies and neutron--skin thickness for
stable neutron rich nuclei are comparable to
the experimental data (see Fig.~1).
For nuclei in the r--process (e.g. $^{132}$Sn--$^{138}$Sn) the
neutron skin has a value of about 0.5\,fm.

With the RMFT we also can determine the proton and neutron density
distribution for the Sn--isotopes. We found acceptable agreement between the
calculated proton--density distributions obtained from RMFT for the stable
Sn--isotopes $^{112}$Sn--$^{124}$Sn with the
experimental data \cite{deV}.
\section{Direct capture for intermediate-- and heavy--mass nuclei}
In this section we want to investigate the importance of DC for intermediate--
and heavy--mass target nuclei in the s-- and r--process.
The DC will be large compared with the CN contribution, if the level
density of the compound nucleus is low, because then there will
be only a few states in the compound nucleus, which can be excited in the
reaction. This is true for example for the radiative
capture reactions on light nuclei cited in the introduction.
For intermediate-- and heavy--mass nuclei this is also the case
if the Q--value of the reaction
{\it as well as\/} the projectile energy are small.

For radiative capture on intermediate-- and heavy--mass target nuclei
induced by charged particles in astrophysical scenarios the projectile
energy is so large that DC can be neglected. For instance, the
DC--contribution of $^{144}$Sm($\alpha$,$\gamma$)$^{148}$Gd is at
least 5 orders of magnitudes lower than the CN--contribution \cite{moh3}.
However, a low level density of the compound nucleus can occur for
radiative capture of neutrons on neutron--magic
and/or neutron rich target nuclei in the s-- or r--process.
\begin{table}
{\bf Table 1}.~DC--calculations of radiative capture
cross sections on neutron--magic target nuclei compared with the experimental
data.
\begin{center}
\begin{tabular}{|c|c|c|}
\hline
Reaction &  $\left<\sigma_{\rm DC} \right>$(25\,keV) [mb] &
$\left<\sigma_{\rm EXP} \right>$(25\,keV) [mb]\\
\hline
$^{48}$Ca(n,$\gamma$)$^{49}$Ca & 0.96 & $1.03\pm0.09$ \cite{kae} \\
$^{86}$Kr(n,$\gamma$)$^{87}$Kr & 0.09 & $3.54\pm0.25$ \cite{bee1} \\
$^{88}$Sr(n,$\gamma$)$^{89}$Sr & 0.26 & 7.0 \cite{bee2} \\
$^{136}$Xe(n,$\gamma$)$^{137}$Xe & 0.16 & $1.05\pm0.09$ \cite{bee1} \\
$^{138}$Ba(n,$\gamma$)$^{139}$Ba & 0.44 & $4.46\pm0.21$ \cite{bee3} \\
$^{208}$Pb(n,$\gamma$)$^{209}$Pb & 0.13 & 0.31 \cite{bee2} \\
\hline
\end{tabular}
\end{center}
\end{table}

In Table 1 we show the results of the DC--contributions
of (n,$\gamma$)--reactions
for some neutron--magic target nuclei with
$N=28$ \cite{krau}, $N=50, 82$ \cite{hub} and
$N=126$ \cite{wuk}
occuring in the s--process.
The DC--calulations were performed using the folding procedure
for the optical and bound--state
potentials as presented in Sect.~2. The average cross
sections $<\sigma>$ at $kT=25$\,keV are compared to the experimental
data. As can be seen the DC gives non--negligible contributions
to the total cross sections. A special case is neutron
capture on the double magic, neutron rich
nucleus $^{48}$Ca, where up to about 1\,MeV
no compound--nucleus levels exist, which can be excited.
Therefore, for this reaction the cross section is
almost given entirely through DC.

As an example for neutron rich nuclei we investigated
the radiative capture on Sn--isotopes up to the r--process
path. The cross sections for such (n,$\gamma$)--reactions are
needed for the description of the freeze--out in the r--process \cite{kl}.
Since
in this case no experimental data is available we
compared the DC-- with the CN--calculations as
obtained from the statistical Hauser--Feshbach (HF)
method. In order to make a meaningful comparison
between the DC-- and CN--reaction mechanism, we
used the same masses, Q--values, spin assignments
and excitation energies and optical potentials
in both calculations.

The densities necessary for the determination of
neutron--nucleus folding potentials (Eq.~2)
involving unstable nuclei cannot be taken any more
from experimental data. We obtained them from the RMFT using the parameter set
NLSH as described in Section 3. The strengths of the
folding potentials were adjusted to reproduce the
same value of the volume integral of 425\,MeV\,fm$^{3}$
as determined from the experimental elastic scattering
data on the stable Sn--isotopes \cite{muh}, \cite{cin}.
In order to obtain experimentally unknown masses
and Q--values a microscopic--macroscopic mass
formula based on the FRDM was used \cite{mol}.
The spin assigments and the excitation energies
of the ground and low--excited states of the odd residual
nuclei $^{133}$Sn--$^{139}$Sn have been taken from
\cite{nym}. For the
neutron spectroscopic factors necessary fo the
DC--calculation a value of unity was assumed.
This assumption should be reasonable for the neutron rich
Sn--isotopes involved in the r--process.

The statistical--model calculations were performed with the code
SMOKER \cite{smo}. Above the highest known state a level density
description based on the
backshifted Fermi gas \cite{gil} is employed, with parameters as
given in \cite{cow}. The imaginary parts of the  optical
potential necessary for the HF--calculations
were taken from \cite{jeu}.
\begin{table}
{\bf Table 2}.~DC-- and HF--calculations of radiative capture
cross sections on neutron rich nuclei.
\begin{center}
\begin{tabular}{|c|c|c|c|}
\hline
Reaction & Q--value [MeV] & $\left<\sigma_{\rm DC} \right>$(30\,keV) [$\mu$b] &
$\left<\sigma_{\rm HF} \right>$(30\,keV) [$\mu$b]\\
\hline
$^{132}$Sn(n,$\gamma$)$^{133}$Sn & 2.581 & 220.7 & 116.4 \\
$^{134}$Sn(n,$\gamma$)$^{135}$Sn & 1.871 & 132.4 & 32.5 \\
$^{136}$Sn(n,$\gamma$)$^{137}$Sn & 1.611 & 101.0 & 18.5 \\
$^{138}$Sn(n,$\gamma$)$^{139}$Sn & 1.230 & 60.4 & 8.4 \\
\hline
\end{tabular}
\end{center}
\end{table}

In Table 2 we show the results of the DC-- compared
to the HF--calculations for the averaged cross
section at 30\,keV on the even target nuclei $^{132}$Sn--$^{138}$Sn.
As can be seen from this
table both cross sections decrease when going
to lower Q--values. However,
the HF--cross section decreases much faster due
to the lower level density of the compound nucleus. For the neutron--capture
on $^{138}$Sn the DC is about 7 times as high
as the CN contribution as obtained with the
HF--method.

Summarizing, we showed that DC is not only
important for radiative capture on light
nuclei, but is also non--negligible for magic and neutron rich
target nuclei. For some double magic and neutron rich
nuclei in the r--process DC is even the dominant reaction mechanism.

Acknowledgments: We want to thank the Hochschuljubil\"aumsfond der Stadt Wien
(project H--82/92) and the DFG for its support. One of us (T.~R.) wants to
thank the
Alexander von Humboldt Foundation.
\end{document}